\documentclass{svmult}
 \pdfoutput=1 

\usepackage[english]{babel}
\usepackage{makeidx} 
\usepackage{graphicx,wrapfig}
\usepackage{multicol}

\usepackage[valuesep=med,unitsep=thin,exponent-product = \cdot]{siunitx}

\usepackage{caption}
\usepackage{subcaption}
\usepackage{rotating}

\usepackage[utf8]{inputenc}
\usepackage{amsmath}
\usepackage{amssymb}
\usepackage{colordvi}
\usepackage{color}
\usepackage{float}
\usepackage{graphicx}
\usepackage{ulem}
\usepackage{wrapfig} 

\usepackage{longtable,lscape}
\usepackage{multirow}
\usepackage{tabularx}

\usepackage{placeins}

\usepackage{ragged2e,array,longtable,lscape}
\usepackage{multirow}
\usepackage{tabularx}
\usepackage{booktabs}

\usepackage{ulem}

\usepackage{placeins}

\usepackage{ulem}
\usepackage{sidecap}

\usepackage{algorithm}
\usepackage{algorithmic}

\usepackage{tikz} 
\usepackage{pgfplots}

\definecolor{DarkGreen}{rgb}{0.133333,0.545,0.1333}
\definecolor{DarkPurple}{rgb}{0.58 , 0.0 ,0.827}

\newdimen\HilbertLastX
\newdimen\HilbertLastY
\newcounter{HilbertOrder}

\def\DrawToNext#1#2{%
   \advance \HilbertLastX by #1
   \advance \HilbertLastY by #2
   \pgfpathlineto{\pgfqpoint{\HilbertLastX}{\HilbertLastY}}
}

\def\Hilbert[#1,#2,#3,#4,#5,#6,#7,#8] {
  \ifnum\value{HilbertOrder} > 0%
     \addtocounter{HilbertOrder}{-1}
     \Hilbert[#5,#6,#7,#8,#1,#2,#3,#4]
     \DrawToNext {#1} {#2}
     \Hilbert[#1,#2,#3,#4,#5,#6,#7,#8]
     \DrawToNext {#5} {#6}
     \Hilbert[#1,#2,#3,#4,#5,#6,#7,#8]
     \DrawToNext {#3} {#4}
     \Hilbert[#7,#8,#5,#6,#3,#4,#1,#2]
     \addtocounter{HilbertOrder}{1}
  \fi
}

\def\hilbert((#1,#2),#3){%
   \advance \HilbertLastX by #1
   \advance \HilbertLastY by #2
   \pgfpathmoveto{\pgfqpoint{\HilbertLastX}{\HilbertLastY}}
     \setcounter{HilbertOrder}{#3}
     \Hilbert[10mm,0mm,-10mm,0mm,0mm,10mm,0mm,-10mm]
     \pgfusepath{stroke}%
  }

\begin{document}
\selectlanguage{\english}
\renewcommand{\baselinestretch}{1.3}\normalsize
\renewcommand{\abstractname}{Abstract}

\title*{
Load balancing strategies for the DSMC simulation of hypersonic flows using HPC
}
\titlerunning{
Hypersonic Flow around a Blunted Cone using HPC
 }
\author{
T.\,Binder\inst{1}\and
S.\,Copplestone\inst{2}\and
A.\,Mirza\inst{1}\and
P.\,Nizenkov\inst{1}\and
P.\,Ortwein\inst{2}\and
M.\,Pfeiffer\inst{1}\and
W.\,Reschke\inst{1}\and
C.-D.\,Munz\inst{2}\and
S.\,Fasoulas\inst{1}
}
\authorrunning{Binder et al.}
\institute{
Institute of Space Systems (IRS), University of Stuttgart, 70569 Stuttgart, Germany
\texttt{fasoulas@irs.uni-stuttgart.de}
\and
Institute of Aerodynamics and Gas Dynamics (IAG), University of Stuttgart, 70569 Stuttgart, Germany
\texttt{munz@iag.uni-stuttgart.de}
}
\maketitle
\setcounter{footnote}{0}
\begin{abstract}
In the context of the validation of PICLas, a kinetic particle suite for the simulation of rarefied, non-equilibrium plasma flows, the biased hypersonic nitrogen flow around a blunted cone was simulated with the Direct Simulation Monte Carlo method. The setup is characterized by a complex flow with strong local gradients and thermal non-equilibrium resulting in a highly inhomogeneous computational load. Especially, the load distribution is of interest, because it allows to exploit the utilized computational resources efficiently. Different load distribution algorithms are investigated and compared within a strong scaling. This investigation of the parallel performance of PICLas is accompanied by simulation results in terms of the velocity magnitude, translational temperature and heat flux, which is compared to experimental measurements.
\end{abstract}

\section{Introduction}\label{sec:introduction}
For the numerical simulation of highly rarefied plasma flows, a fully kinetic modelling of Boltzmann's equation
complemented by Maxwell's equations is necessary. For this purpose a particle codes that combines the PIC (Particle in
Cell) and DSMC (Direct Simulation Monte Carlo) method is developed at IAG (Institute of Aerodynamics and Gas Dynamics)
and IRS (Institute of Space Systems) in recent years \cite{Munz2014}. Particle codes are inherently numerically
expensive and thus are an excellent application for parallel computing. The modelling of the Maxwell-Vlasov equations
(PIC solver) has been described in previous reports~\cite{Stock_etal:2011, ortwein201401, copplestone:HLRS_2016}. In the
present report we focus our attention on the simulation of rarefied, non-equilibrium, neutral gas flows, which are
typical for atmospheric entry conditions at high altitude and are simulated using the DSMC part of the coupled code
PICLas. The inhonogemeous particle distribution throughout the domain leads to strong imbalances. These are reduced
through load balancing for which different load distribution algorithms are investigated.

The physical basis of the coupled solver is the approximation of Boltzmann's equation
\begin{equation}\label{eq:Boltzmann_equation}
\left(\frac{\partial}{\partial t}+\textbf{v}\cdot\nabla+\frac{1}{m^s}
\textcolor{black}{\textbf{F}}
\cdot\nabla_{\textbf{v}}\right)f^s(\textbf{x},\textbf{v},t)=\frac{\partial f}{\partial t}\bigg|_{\mathrm{coll}}~,
\end{equation}
which covers basic particle kinetics, where $f^s(\mathbf{x},\mathbf{v},t)$ is the six-dimensional Particle Distribution Function (PDF) in phase-space for each species $s$ with mass $m$. It describes the amount of particles per unit volume, which are found at a certain point $(\vec{x},\vec{v})$ in phase-space and time $t$. The left hand side of~\eqref{eq:Boltzmann_equation}, where $\textbf{F}$ is an external force field, is solved using a deterministic Particle-in-Cell~\cite{hockney198801} method, while the right hand side, where the collision integral $\frac{\partial f}{\partial t}\big|_{Coll}$ accounts for all particle collisions in the system, is solved by applying the non-deterministic DSMC~\cite{bird199401} method.

The PDF is approximated by summing up a certain number of weighted particles $\N_{\mathrm{part}}$ and is given by
\begin{equation*}
f^s(\textbf{x},\textbf{v},t)\approx\sum_{n=1}^{N_\mathrm{part}} w_\mathrm{n} \delta\left(\textbf{x}-\textbf{x}_\mathrm{n}\right)\delta\left(\textbf{v}-\textbf{v}_\mathrm{n}\right),
\end{equation*}
where the $\delta$-function is applied to position and velocity space, separately, and the particle weighting factor $w_\mathrm{n}=N_{\text{phy}}/N_{\text{sim}}$ is used to describe the ratio of simulated to physical particles.

The DSMC method is briefly reviewed in Section~\ref{sec:DSMC}. In Section~\ref{sec:bluntedCone}, the numerical setup and results of the simulation of the flow around a 70$^\circ$ blunted cone 
geometry are presented. The load-distribution algorithms and the parallel performance of the DSMC code are investigated in detail in Section~\ref{sec:parallel}, followed by a summary and conclusion in Section~\ref{sec:conclusions}.

\section{DSMC Solver}\label{sec:DSMC}\label{sec:dsmc}
The DSMC method approximates the right hand side of Eq.~\eqref{eq:Boltzmann_equation} by modelling binary particle collisions in a probabilistic and transient manner. The main idea of the DSMC method is the non-deterministic, statistical calculation of changes in particle velocity utilizing random numbers in a collision process. Additionally, chemical reactions may occur in such collision events. The primordial concept of DSMC was developed by Bird~\cite{bird199401} and is commonly applied to the simulation of 
rarefied and neutral gas flows. The collision operator in Eq.~\eqref{eq:Boltzmann_equation} is given by
\begin{equation}
\begin{split}
\frac{\partial f}{\partial t}\big|_{\mathrm{coll}}=\qquad\qquad\qquad\qquad\qquad\qquad\qquad\qquad\qquad\qquad\qquad\qquad\qquad\qquad\\
\int W(\textbf{v}_1,\textbf{v}_2,\textbf{v}_3,\textbf{v}_4)\lbrace
f(\textbf{x},\textbf{v}_1,t)
f(\textbf{x},\textbf{v}_2,t)-
f(\textbf{x},\textbf{v}_3,t)
f(\textbf{x},\textbf{v}_4,t)
\rbrace
d\textbf{v}_1 d\textbf{v}_2 d\textbf{v}_3~,
\end{split}
\end{equation}
where $W$ represents the probability per unit time in which two particles collide and change their velocities from $\textbf{v}_1$ and $\textbf{v}_2$ to $\textbf{v}_3$ and $\textbf{v}_4$, respectively. However, the DSMC method does not solve this collision integral directly, but rather applies a phenomenological approach to the collision process of simulation particles in a statistical framework.

\begin{figure}[ht] 
  \centering 
  \includegraphics[width=0.8\textwidth]{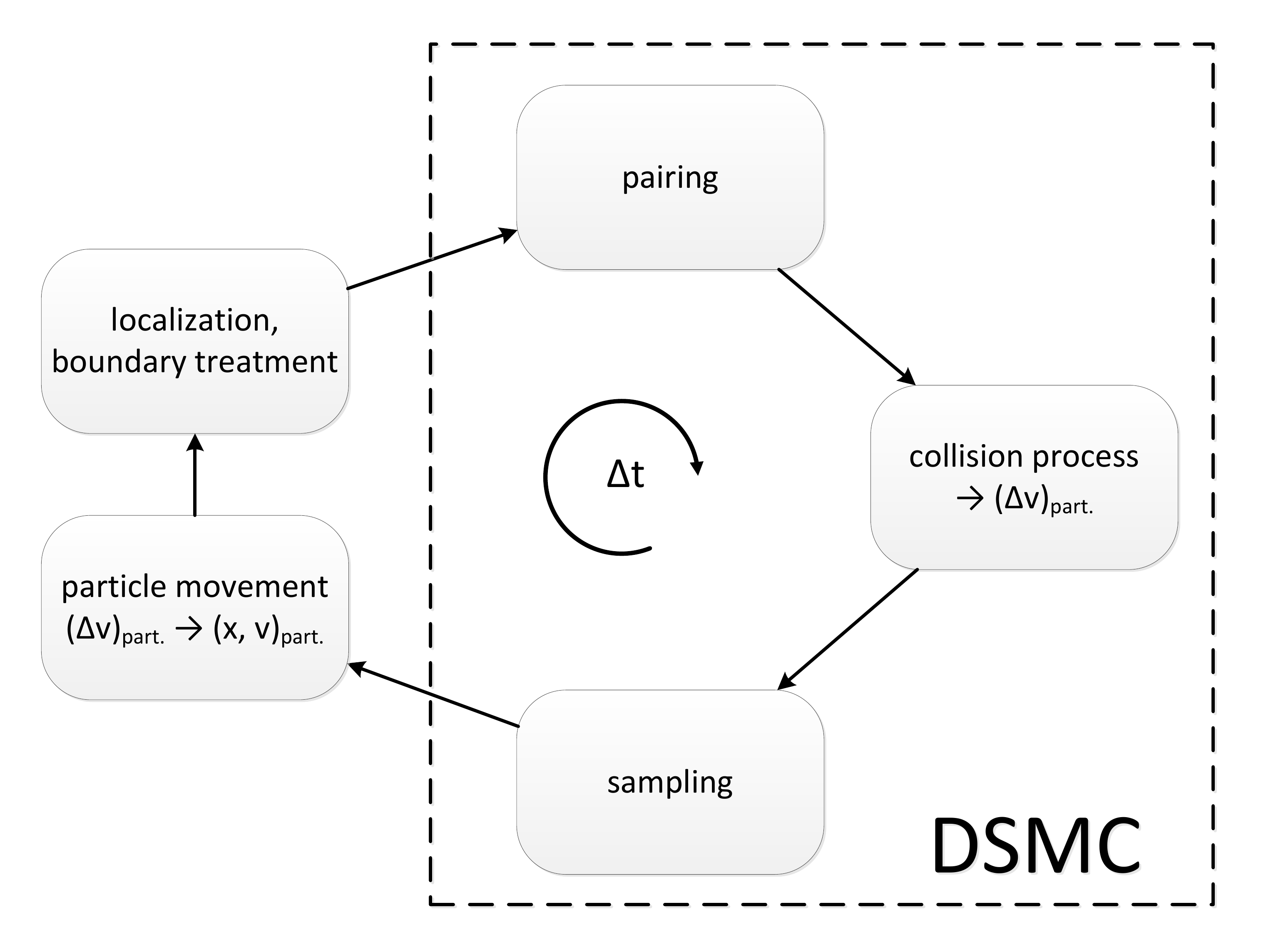}
  \caption{Schematic of the standard DSMC method.}
  \label{fig:DSMC_cycle}
\end{figure}

A single standard DSMC time step is depicted schematically in Fig.~\ref{fig:DSMC_cycle}. 
First, a particle pair for the collision process is found by examining each cell and 
applying a nearest neighbour search with an octree based pre-sorting. An alternative method is the random pairing of all particles in each cell, but with additional restrictions to the cell size. The collision probability is modelled by choosing a cross section for each particle species using microscopic considerations. As with the PIC method, macro particles are simulated instead of real particles to reduce computational effort. The collision probability of two particles, $1$ and $2$, is determined by 
methods found in~\cite{bird199401,baganoff1990}, which yields
\begin{equation}
P_{12}=\frac{N_{p,1}N_{p,2}}{1+\delta_{12}}w\frac{\Delta t}{V_cS_{12}}(\sigma_{12}g_{12})~,
\end{equation}
where $\delta_{12}$ is the Kronecker delta, $V_\mathrm{c}$ the cell volume, $\Delta t$ the time step, $\sigma$ the cross section, $S_{12}$ the number of particle pairs of species $1$ and $2$ in $V_\mathrm{c}$
and $g$ the relative velocity between the two particles considered. This probability is compared to a pseudo random number $R\in [0,1)$ and if $R<P_{12}$, the collision occurs, otherwise it does not. Subsequent events such as chemical reactions or relaxation processes are computed in the same manner, but using additional probabilities. This may change the internal energy of particles, i.e. their rotational, vibrational energy and electronic excitation. Chemical reactions are modelled via the Arrhenius law or quantum-kinetic considerations, which lead to dissociation, recombination, exchange reactions or ionization. Macroscopic properties like temperature or density are calculated by sampling particle positions and velocities over time within each cell.

A major requirement for a physical DSMC simulation is the ratio of the mean collision separation distance to the mean free path in each cell 
\begin{equation}
\frac{l_\mathrm{mcs}}{\lambda} \overset{!}{<} 1.
\end{equation}
The former represents the distance of two simulation particles that perform a collision, while the latter is a function of the gas density. The ratio can be modified by the weighting factor $w_\mathrm{n}$ as introduced in Section \ref{sec:introduction}, which then directly depends on the local number density
\begin{equation}
w < \frac{1}{\left(\sqrt{2}\pi d_\mathrm{ref}^2 n^{2/3}\right)^3},
\end{equation}
where $d_\mathrm{ref}$ is a species-specific reference diameter.

\section{Test Case: $70^\circ$ Blunted Cone}\label{sec:bluntedCone}

\begin{figure}[tb]
\centering
\begin{tikzpicture}
\begin{axis}[%
      extra description/.code={%
        font=\small
        \node[anchor=north east, fill=white] at (190,80) {%
		\begin{tabular}{@{}cc@{}}
		\hline\noalign{\smallskip}
		\# & $S/R_\mathrm{n}$ $\left[-\right]$\\
		\noalign{\smallskip}\hline\noalign{\smallskip}
		1 & 0.00 \\
		2 & 0.52 \\
		3 & 1.04 \\
		4 & 1.56 \\
		5 & 2.68 \\
		6 & 3.32 \\
		7 & 5.06 \\
		8 & 6.50 \\
		9  & 7.94 \\
		\noalign{\smallskip}\hline
		\end{tabular}
        };
        \node[anchor=north east, fill=white] at (-130,80) {%
\small
      		\begin{tabular}{@{}cc@{}}
      		\hline\noalign{\smallskip}
      		 & [\si{\milli\meter}] \\
      		\noalign{\smallskip}\hline\noalign{\smallskip}
      		$R_\mathrm{b}$ & 25.0 \\
      		$R_\mathrm{c}$ & 1.25 \\
      		$R_\mathrm{j}$ & 2.08 \\
      		$R_\mathrm{n}$ & 12.5 \\
      		$R_\mathrm{s}$ & 6.25 \\
      		\noalign{\smallskip}\hline
      		\end{tabular}
        };
      },
      hide axis
    ]
    \pgftext{\includegraphics[width=0.65\linewidth]{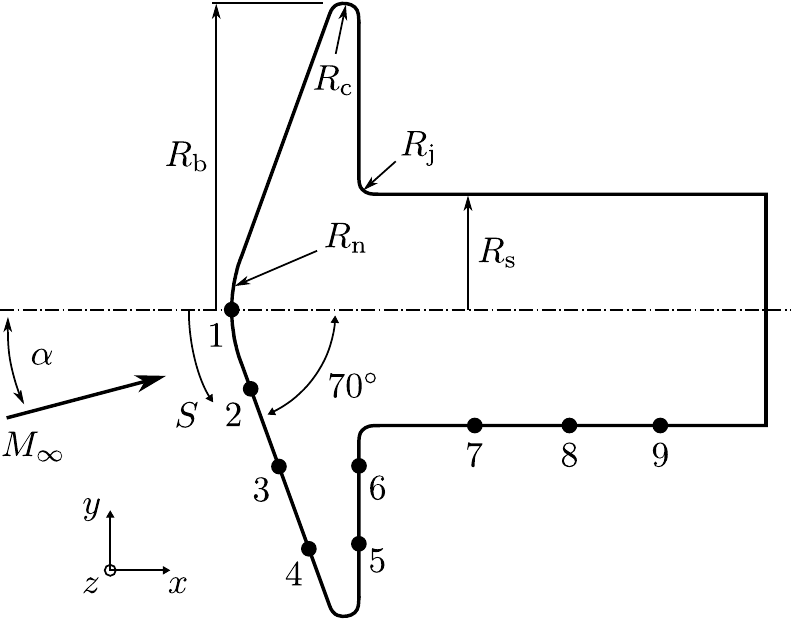}}
    \end{axis}
\end{tikzpicture}
\caption{Geometry of the $70^\circ$ blunted cone test case.}\label{fig:70degCone_geometry}
\end{figure}

A popular validation case for rarefied gas flows is the wind tunnel test of the $70^\circ$ blunted cone in a diatomic nitrogen flow at a Mach number of $M=20$~\cite{Allegre1997}. The geometry of the model is depicted in Fig.~\ref{fig:70degCone_geometry}. Positions of the heat flux measurements are depicted by the numbers 1-9. While the experiments were conducted at different rarefaction levels and angles of attack, the case denoted by Set 2 and $\alpha=30^\circ$ is used for the investigation. The free-stream conditions and simulation parameters are given in Table~\ref{tab:70degCone_freestream}. Half of the fluid domain was simulated to exploit the symmetry in the $xy$-plane.

\begin{table}[tb]
	\caption{Free-stream conditions of the $70^\circ$ blunted cone test case.}\label{tab:70degCone_freestream}
	\centering
	\begin{tabular}[tb]{@{}lcccccc@{}}
		\hline
		Case & $|\vec{v}_\infty|$ $\left[\si{\metre\per\second}\right]$ &  $T_\infty$ [\si{\kelvin}] &  $n_\infty$ [\si{\per\cubic\metre}] & $\Delta t$ $\left[\si{\second}\right]$ &  $w$ $[-]$ & $N_\mathrm{part}$ $\left[-\right]$ \\
		\hline
		Set 2  & 1502.4 & 13.58 & \num{1.115E+21} & \num{5e-8} & \num{2e10} & \num{2.84e+07} \\
		\hline
	\end{tabular}
\end{table}

\begin{figure}[p]
	\centering
\begin{tikzpicture}
\begin{axis}[
	hide axis,
	axis equal image,
	width=14cm,
	xmin=0,
	xmax=1268,
	ymin=0,
	ymax=1699,
	colormap={rgb}{rgb255=(255,255,255) rgb255=(0,0,255) rgb255=(0,255,255) rgb255=(0,255,0) rgb255=(255,255,0) rgb255=(255,0,0) rgb255=(255,0,255)},
	colorbar right,
	point meta min=15,
	point meta max=1250,
	colorbar style={
			scaled y ticks = false,
			ytick={15,250,500,750,1000,1250},
			height=2.5cm,
			at={(-0.1,0.525)},
			anchor=south east,
			title={$T$ $[\si{\kelvin}]$},
		},
		colorbar/draw/.append code={
		\begin{axis}[
			colormap={rgb}{rgb255=(0,0,255) rgb255=(0,255,255) rgb255=(0,255,0) rgb255=(255,255,0) rgb255=(255,0,0)},
			colorbar right,
			point meta min=0,
			point meta max=1500,
            every colorbar,
            at={(-0.1,0.125)},
			ytick={0,500,1000,1500},
            anchor=south east,
            width=0.5cm,
            colorbar shift,
            colorbar=false,
			title={$v$ $[\si{\metre\per\second}]$},
			]
			\pgfkeysvalueof{/pgfplots/colorbar addplot}
		\end{axis}
	    }
		]	
	\addplot graphics [xmin=0,xmax=1268,ymin=0,ymax=1699] {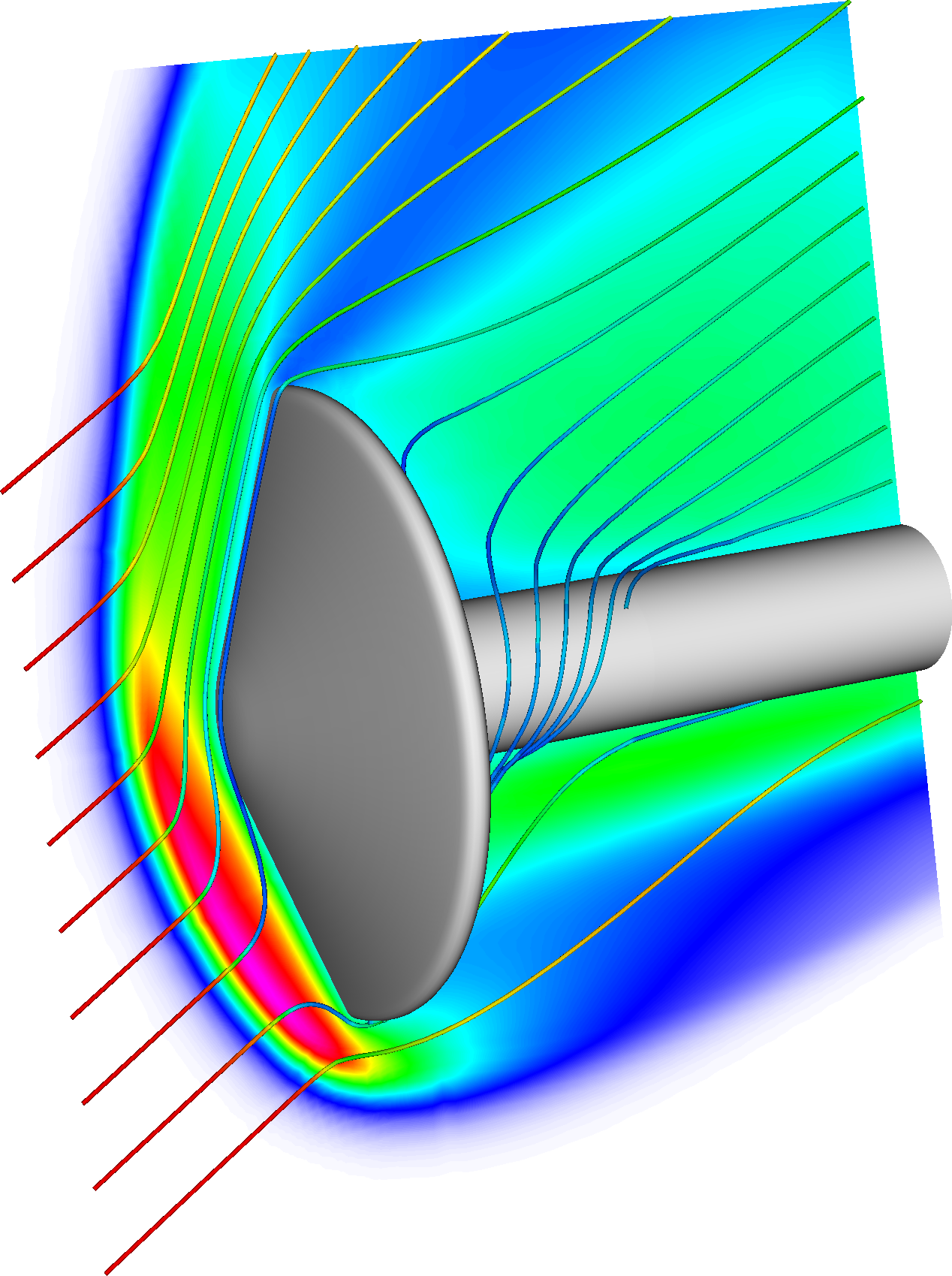};
\end{axis}
\end{tikzpicture}
	\caption{Exemplary simulation result: Translational temperature in the symmetry plane and velocity streamlines.}\label{fig:70degCone_Set2_3D}
\end{figure}

\begin{figure}[p]
	\centering
\begin{tikzpicture}
\tikzset{
small dot/.style={fill=black,circle,scale=0.3},
axis/.style={->, >=stealth'},
every pin edge/.style={draw=white},
every pin/.style={font=\footnotesize}
}
\begin{semilogyaxis}[
width=8cm,
height=5.75cm,
xlabel={$S/R_\mathrm{n}$ $[-]$},
ylabel={Heat flux $q_\mathrm{w}$ $\left[\si{\kilo\watt\per\square\metre}\right]$},
ymin=0.01,
ymax=100,
xmin=-0.3,
xmax=8.3,
xtick={0,1,...,8},
xlabel shift=-2 pt,
legend style={
	font=\footnotesize,
	legend cell align=left,
	at={(0.95,0.075)},
	anchor=south east,
			fill=white,
			draw=black}
]

\addplot+[
    nodes near coords,
    point meta=explicit symbolic,
    nodes near coords align={anchor=south, above, yshift=3mm},
    color=black, mark=x, only marks, font=\footnotesize]  table [col sep=comma, x=position, y=alpha_30, meta=thermocouple] {Allegre1997_set2_heatflux.csv};

\addlegendentry{Experiment}

\addplot[color=black]  table [col sep=comma, x=SoverR, y=Heatflux_kW] {70degCone_Set2_alpha30_07ms.csv};
\addlegendentry{PICLas}

\end{semilogyaxis}
\end{tikzpicture}
	\caption{Comparison of measured and calculated heat flux.}\label{fig:70degCone_Set2_Heatlfux}
\end{figure}
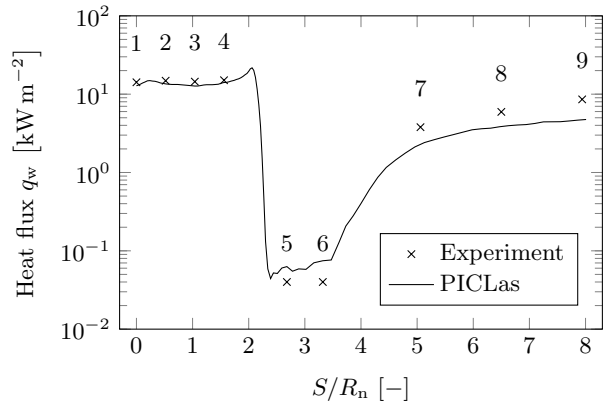

An exemplary simulation result is shown in Fig.~\ref{fig:70degCone_Set2_3D}. Here, the translational temperature in the symmetry plane and the velocity streamlines are shown. The simulation results are compared to the experimental measurements in terms of the heat flux in Fig.~\ref{fig:70degCone_Set2_Heatlfux}. Overall good agreement can be observed for the first four thermocouples, where the error is below $10\%$ and within experimental uncertainty~\cite{Allegre1997}. The agreement on the sting deteriorates for thermocouples further downstream to error values of up to $45\%$.

\section{Parallelization of the DSMC Method}   \label{sec:parallel}
\subsection{Load Computation and Distribution}
\label{sec:load_distribution}
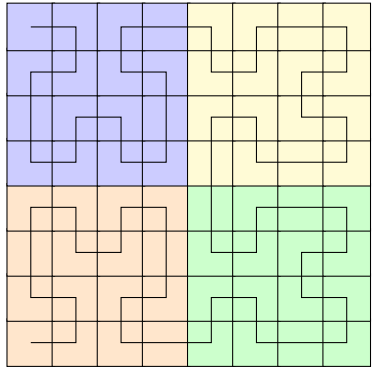
\begin{figure}
  \def\SizeW{0.44\textwidth}
    \centering
\begin{tikzpicture}[scale=0.6,every node/.style={transform shape}]
 \foreach \i in {1,...,16}
  {
      \pgfmathtruncatemacro{\y}{(\i-1 ) / 4};
      \pgfmathtruncatemacro{\x}{\i -1 - 4 * \y};
      \pgfmathtruncatemacro{\label}{\x +7 * (7 - \y)};
      \node[rectangle,draw=black,fill=white!80!orange,minimum size=30] () at (\x,\y) {};
  }
  \foreach \i in {17,...,32}
  {
      \pgfmathtruncatemacro{\y}{(\i-1 ) / 4};
      \pgfmathtruncatemacro{\x}{\i -1 - 4 * \y};
      \pgfmathtruncatemacro{\label}{\x +7 * (7 - \y)};
      \node[rectangle,draw=black,fill=white!80!blue,minimum size=30] () at (\x,\y) {};
  }
  \foreach \i in {1,...,16}
  {
      \pgfmathtruncatemacro{\y}{(\i-1 ) / 4};
      \pgfmathtruncatemacro{\x}{\i +3 - 4 * \y};
      \pgfmathtruncatemacro{\label}{\x +7 * (7 - \y)};
      \node[rectangle,draw=black,fill=white!80!green,minimum size=30] () at (\x,\y) {};
  }
  \foreach \i in {17,...,32}
  {
      \pgfmathtruncatemacro{\y}{(\i-1 ) / 4};
      \pgfmathtruncatemacro{\x}{\i +3 - 4 * \y};
      \pgfmathtruncatemacro{\label}{\x +7 * (7 - \y)};
      \node[rectangle,draw=black,fill=white!80!yellow,minimum size=30] () at (\x,\y) {};
  }
 \hilbert((0mm,0mm),3)
\end{tikzpicture}
    \caption{Domain decomposition for homogeneous load distribution.}
    \label{fig:hilbert-domain}
\end{figure}

The code framework of PICLas utilizes implementations of the MPI 2.0 standard for parallelization.
Load distribution between the MPI processes is a crucial step. A domain decomposition by grid elements was chosen as strategy. 
In a preprocessing step, all elements within the computational domain are sorted along a Hilbert 
curve due to its clustering property \cite{moon200101}. Then, each MPI process receives a certain 
segment of the space filling curve (SFC). 
To illustrate an optimal load balance scenario, a simplified grid is considered that consists of 
$8 \times 8 = 64$ elements, which are ordered along a SFC. Fig.~\ref{fig:hilbert-domain} depicts the decomposition of the grid into four regions, each corresponding to an individual MPI process when the number of processes is $N_{p}=4$.
For inhomogeneous particle distributions or elements of significantly different size, the load has to be assigned carefully. In the DSMC method, the computational costs $L$ of each grid element is assumed to be linearly dependent on the contained particle number.
In an optimally balanced case, each process receives approximately the average load.

Offset elements (i.e., an index $I$ along the SFC) define the assigned segment of a process. When the SFC describes the interval of $[1,N_{elem}]$, the segment of each process $p$ is defined by $[I(p)+1,I(p+1)]$ with $I(N_{p}+1)=N_{elem}$.
Thus, the total load assigned to a single process results in:
\begin{equation}
L_{tot}^{p}=\sum_{i=I(p)+1}^{I(p+1)} L_{i}
\end{equation}

The main goal of a proper load distribution is to minimize the idle time of waiting processes, i.e., the maximum of all total, process-specific loads $L_{tot}^{p}$ needs to be minimized. To achieve that, several distribution methods are implemented in PICLas.

\paragraph{Distribution by elements}
Assuming a homogeneous particle population, a distribution only by elements is favourable, i.e., $L_{elem}=$const. This can be achieved by dividing the number elements into:
\begin{equation}
N_{Elems}=N_{p}\cdot A+B,\quad A=\left \lceil{\frac{N_{Elems}}{N_{p}}}\right \rceil,\quad B=N_{Elems}~\mathrm{mod}~N_{p}
\end{equation}
Based on this, each process receives $A$ elements and the first $B$ processes an additional one, which can be calculated in a straightforward manner by:

\begin{algorithm}
\caption{Distribution by elements}
\begin{algorithmic}
\STATE $i_{p}\gets 1$
\WHILE{$i_{p}\leq N_{p}$}
\STATE $I(i_{p})\gets A\cdot(i_{p}-1)+\mathrm{min}(i_{p}-1,B)$
\STATE $i_{p}\gets i_{p}+1$
\ENDWHILE
\STATE $I(N_{p}+1)\gets N_{elem}$
\end{algorithmic}
\end{algorithm}

\paragraph{Simple load balance}
The previous method is, however, not applicable if the elements have different loads, since a subdivision in element number does not necessarily correspond in the same fraction of total load.
Therefore, while looping through the processes along the SFC, each process receives in our ``simple'' balance scheme an iteratively increasing segment until the so far gathered load is equal or greater than the ideal fraction.
To ensure that the following processes receive at least one element each, the respective number of assignable elements must be reduced. The algorithm follows as:

\begin{algorithm}
\caption{Simple load balance}
\begin{algorithmic}
\STATE $L_{tot}\gets 0$
\STATE $i_{elem}\gets 1$
\STATE $i_{p}\gets 1$
\WHILE{$i_{p}\leq N_{p}$}
\STATE $I(i_{p})\gets i_{elem} - 1$
\STATE $j\gets i_{elem}$
\WHILE{$j\leq N_{elem} - N_{p} + i_{p} \quad\land\quad L_{tot} < \frac{i_{p}}{N_{p}}\cdot \sum_{k=1}^{N_{elem}} L_{k}$}
\STATE $L_{tot}\gets L_{tot} + L_{j}$
\STATE $j\gets j+1$
\ENDWHILE
\STATE $i_{elem}\gets j + 1$
\STATE $i_{p}\gets i_{p}+1$
\ENDWHILE
\end{algorithmic}
\end{algorithm}

\paragraph{``Combing'' algorithm}
The ``simple''  algorithm ensures a very smooth load distribution for large element numbers, since the ideal, current fraction can be achieved well by the iterative adding of elements.
However, if there exist elements with much higher loads than most of the remaining ones, the load distribution method fails.
For this, we developed a smoothing algorithm, that ``combs'' the offset elements along the SFC iteratively from the beginning towards the end.
Here, just the main characteristics of the method should be briefly described:
\begin{itemize}
\item The initial load distribution follows, i.e., from the ``simple'' balance method.
\item A large number of different distributions is evaluated in terms of the maximum process-total load $\mathrm{max}(L_{tot}^{p})$, the one with the minimum value is chosen as final solution.
\item If the maximum $L_{tot}^{p}$ belongs to a process $p$ with a greater SFC-index than the minimum one (maximum is ``right'' of the minimum), all offset elements are shifted accordingly to the left.
\item Maxima are smoothed to the right, i.e., small $L_{tot}^{p}$-intervals are increased by shifting elements from maxima to minima.
\item If the resultant optimum distribution was already reached before, elements are shifted from the last process all towards the first one.
\end{itemize}

\subsection{Scaling performance of PICLas}

For the test of parallelization, multiple simulations were run for a simulation time of \SI{1e-4}{\second}, corresponding to \num{2000} iterations. The speed-up between \num{720} and \num{5760} cores was calculated by
\begin{equation}
S_N=\frac{t_{720}}{t_N}.
\end{equation}
The respective parallel efficiency was determined by
\begin{equation}
\eta_N=\frac{720\cdot t_{720}}{N \cdot t_N},
\end{equation}
where $t_{720}$ and $t_N$ is the computational time using 720 and $N$ cores, respectively.

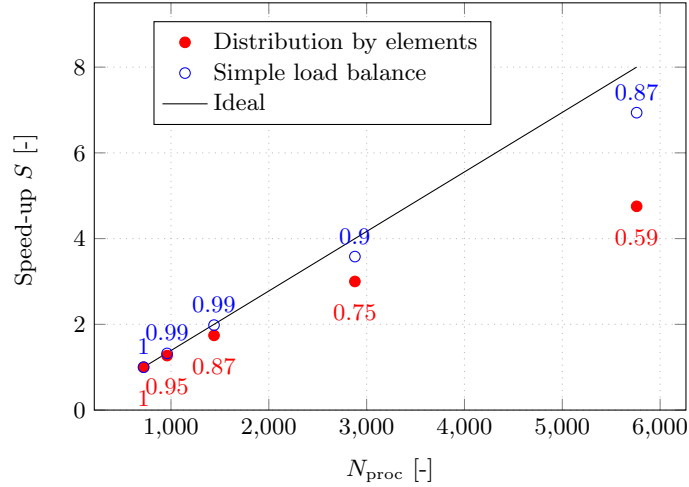
\begin{figure}[htb]
  \def\SizeW{0.8\textwidth}
    \centering
\begin{tikzpicture}
\begin{axis}[
font=\footnotesize,
width=0.8\textwidth,
height=7cm,
grid = minor,
grid=both,
grid style={dotted},
ymin = 0,
ymax = 9.5,
xlabel={$N_{\mathrm{proc}}$~[-]},
ylabel={Speed-up $S$~[-]},
ylabel style={at={(0.04,0.7)}, anchor=east},
legend cell align=left,
legend pos= south east,
legend style={at={(0.1,0.7)},anchor=south west}
] 
\addplot[mark=*, visualization depends on=720*366/(\thisrowno{0}*\thisrowno{1}) \as \labela, color=red, only marks, nodes near coords=\pgfmathprintnumber{\labela}, 
every node near coord/.append style={
shift={(axis direction cs:0,-1.1)}
}
] 
table[header=false,x expr=\thisrowno{0},y expr=366/\thisrowno{1}, col sep=comma]
{./wo_scal.csv};
\addlegendentry{Distribution by elements}

\addplot[mark=o, visualization depends on=720*111/(\thisrowno{0}*\thisrowno{1}) \as \labela, color=blue, only marks, nodes near coords=\pgfmathprintnumber{\labela}, 
every node near coord/.append style={
shift={(axis direction cs:0, 0.1)}
}
] 
table[header=false,x expr=\thisrowno{0},y expr=111/\thisrowno{1}, col sep=comma]
{./case0.csv};
\addlegendentry{Simple load balance}

\addplot[mark=none, color=black] table[header=false,x expr=\thisrowno{0},y expr=\thisrowno{0}/720, col sep=comma]
{./wo_scal.csv};
\addlegendentry{Ideal}

\end{axis}
\end{tikzpicture}
\caption{Parallel performance of the double cone test case between 720 and 5670 cores. 
Speed-up $S$ with labelled parallel efficiency $\eta$.}\label{fig:strongscale}
\end{figure}

Fig.~\ref{fig:strongscale} shows the speed-up over the number of utilized nodes and the respective parallel efficiency as a label.
The case without actual load balance (distribution by elements) is compared together with the distribution method by paticle number per element against the ideal scaling behavior.
The ``Combing'' algorithm resulted into the same performace values as the ``simple'' balance method, therefore, only the the latter one is displayed.
The speed-up decreases with an increasing number of cores due to the more frequent communication between MPI processes. Nevertheless, a parallel efficiency of $\eta=0.87$ can be achieved using \num{5760} cores for the blunted cone test case. 

\section{Summary and Conclusions}\label{sec:conclusions}
The hypersonic flow around a $70^\circ$ blunted cone was simulated with the Direct Simulation Monte Carlo method. The case features complex flow phenomena such as a detached compression shock in front and rarefied gas flow in the wake of the heat shield. A comparison of the experimentally measured heat flux yielded good agreement with the simulation results.
The test case was utilized to perform a strong scaling of the DSMC implementation of PICLas. With regard to the computational duration on \num{720} cores, a parallel efficiency of $99\%$ to $87\%$ could be achieved for \num{1440} and \num{5760} cores, respectively. The decrease in parallel efficiency can be explained by an increasing MPI communication effort. Currently, the implementation of cpu-time measurements into PICLas is investigated for calculating the element loads directly instead of a simple weighting of particle number, which will be focus of future reports.

\section{Acknowledgements}
\label{sec:acknowledgements}
We gratefully acknowledge the Deutsche Forschungsgemeinschaft (DFG) for funding within the projects "Kinetic Algorithms for the Maxwell-Boltzmann System and the Simulation of Magnetospheric Propulsion Systems" and "Coupled PIC-DSMC-Simulation of Laser Driven Ablative Gas Expansions". The latter being a sub project of the Collaborative Research Center (SFB) 716 at the University of Stuttgart. The authors also wish to thank the Landesgraduier\-tenf\"{o}rderung Baden-W\"{u}rttemberg for supporting the research. Computational resources have been provided by the H\"ochst\-leistungs\-rechen\-zentrum Stuttgart (HLRS).

\bibliographystyle{plain}
\bibliography{./loc_IMPD.bib}

\end{document}